\documentstyle[aps,preprint]{revtex}

\def\ba{\begin{eqnarray}}
\def\ea{\end{eqnarray}}
\begin{document}
\bibliographystyle{sort}
\begin{titlepage}{
\vspace{4cm}
\begin{center}
{ \LARGE Landauer and Thouless Conductance: a Band Random Matrix Approach}

\vspace{0.8cm} 

{\large Giulio Casati$^{(a,b)}$ 
, Italo Guarneri$^{(a,c)}$, Giulio Maspero }

\vspace{0.5cm}

{ \em
International Center for the Study of Dynamical Systems,
University of Milan at Como, via Lucini 3, 22100 
Como, Italy 

$^{(a)}$ Istituto Nazionale di Fisica della Materia, Unit\'a di Milano,
 via Celoria 16, 20133, Milan, Italy 

$^{(b)}$ I.N.F.N., Sezione di Milano, via Celoria 16, 20133, Milan, Italy 

$^{(c)}$ I.N.F.N., Sezione di Pavia, via Bassi 6, 27100, Pavia, Italy }
\end{center}
\vspace{1.5cm}
{

 
\vspace{1cm}

{ \large Section: Condensed Matter}

\vspace{1cm}

G.Maspero

e-mail: maspero@mvcomo.fis.unico.it

fax: ++39 31 326230
}}
\end{titlepage}

\begin{abstract}
We numerically analyze the transmission through a thin disordered 
wire of finite length attached to perfect leads, by making use of banded
random Hamiltonian matrices.  We compare  
the Landauer 
and the Thouless conductances, and find that they are proportional 
to each other in 
the diffusive regime, while in the localized regime the Landauer conductance 
is approximately proportional to the square of the Thouless one.
Fluctuations of the Landauer conductance were also numerically computed;
they are shown to slowly approach the theoretically predicted value.  
\end{abstract}


\newpage


Two main theoretical approaches to the study of electronic conduction 
in disordered wires at zero temperature have been developed, starting from 
ideas originally due to Thouless and Landauer, respectively. In the Thouless 
approach, the wire is regarded as a closed system, and its 
conductance is defined through the sensitivity of eigenvalues to changes of 
the boundary conditions \cite{Thouless72,Thouless74}.
In the Landauer approach\cite{landauer57,landauer70}  the wire is an open 
system, 
connected to perfect conductors, and its  
conductance is related to the transparency of the 
wire to electronic waves, via a formula, that can be derived from linear 
response theory\cite{fish}. 
 In the metallic regime, 
the conductance is given by the 
Thouless formula, which can in turn be derived from the 
Kubo-Greenwood formula 
supplemented by certain random-matrix theoretical assumption \cite{akkermans}. 
The relation between the above two approaches is a subtle theoretical problem.

 In this Letter we present a
comparison of the Thouless and the Landauer conductance, on a model which 
allows for an efficient numerical computation of both.  In this model, the 
wire is described by a 'Band Random Matrix' (BRM), that is,   
a real symmetric 
matrix $H$ of rank $L$ such that matrix elements
$h_{ij} \neq 0$ only for $\left| i-j 
\right| \leq b+1$, where $b$ is the halfwidth of the band. 
We chose Gaussian distribution for all non-zero elements, with 
variance ${\alpha^2 \over 2}$ for those on the  side diagonals
 and ${\alpha^2}$ for those on the main diagonal . Strictly speaking, BRMs of 
this kind describe 1D disordered wires with long-range hopping; however, they 
are also deemed to catch most of the general features of quasi-one dimensional 
models \cite{fyod}.
 
In the Landauer approach the wire is connected 
on both sides to ideal conductors, or "leads".
In our model the electron 
dynamics in the leads is described by  
two semi-infinite BRMs with the same bandwidth $b$, with  all matrix elements 
inside the band  equal to $1$. The Schr\"{o}dinger equation is
\ba \label{hamil}
 \sum_{k=-(b+1)}^{b+1} {\cal{H}}_{i,i+k} u_{i+k} = E u_{i}
\ea
where ${\cal{H}}_{i,j}$ are the non-zero elements of the Hamiltonian 
$\cal{H}$, which is an infinite band matrix. 
Elements  ${\cal{H}}_{i,j}$ are the same as above defined  ${{h}}_{i,j}$
for $1 \leq i,j \leq L$ and are equal to $1$ otherwise.
In this way, one can write
${\cal{H}} = {\cal{H}}_{0}+V$, where ${\cal{H}}_0$ is an infinite band matrix,
with constant unit elements inside the band, which describes free motion,
and $V$ is a scattering potential.
The original BRM ${{H}}$ is just the projection of the Hamiltonian
${\cal{H}}$ on the subspace spanned by states $\left| i \right> $, $1 \leq
i \leq L$.

Free waves  (which are eigenfunctions of ${\cal{H}}_0$) 
 are of the form   
$u_m={1 \over{ \sqrt{ 2 \pi } }} e^{i k m}$, 
with the wave number $k$ obeying the dispersion law
\ba \label{dispr}
{\sin{({{2 b +1}\over 2} k)} \over {\sin{( {k\over2})}}} = E.
\ea

The left hand side of (\ref{dispr}) is a periodic function of $k$ of period
$2\pi$, which takes the value $2b+1$ when its  
argument is a multiple of $2\pi$.
For $-1 < E < +1$
Eq.(\ref{dispr})
 has  $2b$ solutions, which can be found 
analytically for $E=1$, $E=0$ and $E=-1$ or
by numerical methods at any other value of $E$.
For example, they are
 $k_j= {2\pi j\over{2b+1}}$ with $j=1, \ldots, 2b$
for $E=0$ and
$k_s={2\pi s\over{b}}$ with $s=1, \ldots, b-1$ and 
$k_t={2\pi t+\pi \over {b+1}}$
with $t=0, \ldots, b$ for $E=1$. The various allowed values of the wave-number 
$k$ at a given energy $E$ correspond to different scattering channels and
in the range $ 1 \geq E \geq -1$ all scattering channels are open. 
We  thus have
a multi-channel scattering problem with exactly $b$ ingoing and $b$ outgoing 
channels, which, if $1<<b<<L$ , formally reproduces 
the case of a quasi-one dimensional wire.

The scattering properties of the conductor are
described by a unitary scattering matrix $S$, that relates incoming and
outgoing amplitudes, $I_{L(R)}$ and $O_{L(R)}$:
\[
S    \left(
\matrix{I_L \cr I_R} \right) 
   =    \left( \matrix{O_L \cr O_R} \right)
\]
where the subscripts $L$ and $R$ stand for the left and the right lead.
The $S$ matrix can be written as
\ba \label{S}
S=\left[
\matrix{r & t \cr
t' & r'} \right]
\ea
where $t$ and $t'$ are the transmission sub-matrices in the 
two opposite directions, and $r$ and $r'$ are the reflection
sub-matrices.

The Landauer conductance is \cite{Buttiger}
\ba \label {land-for}
{\cal{G}} ={\frac{e^2}{2 \pi \hbar}} \sum_{ij} {\left| t_{ij} \right|^2},
\ea
where $t_{ij}$ are the elements of the $t$ sub-matrix in (\ref{S}).
We have computed
the S-matrix by standard methods, namely 
by numerically solving the Lippman-Schwinger 
equation
\ba
 u^\pm-G_0^\pm\cdot V u^\pm = u 
\ea
where $u$ are free eigenfunctions, $u^\pm$ interacting ones.
The free Green function, $G_0^\pm =(E-H_0{\pm}i \epsilon)
^{-1}_{\epsilon=0}$, has matrix elements 
                  
                  \begin{eqnarray} 
                  (G_0^\pm)_{n,m} & =\langle 
                  n|(E-H_0{\pm}i \epsilon)^{-1}|m\rangle 
                  \big|_{\epsilon=0} \nonumber \\
                  & = {1\over {2 \pi}} \int_0^{2 \pi} dk 
                  {e^{i(m-n) k}\over{E{\pm}i \epsilon
                  -{\sin{({{2 b +1}\over 2} k)} \over {\sin{( {k\over2})}}}}}
                  \Bigg|_{\epsilon=0}     \nonumber
                  \end{eqnarray}
which were numerically computed by exploiting the Residues Theorem.
Finally, the scattering matrix was computed via
 the standard formula:

\ba \label{scat}
S_{i j}=\delta_{i j}- 
2 \pi i \sqrt{{dk_i \over dH}} \Bigg|_{H=E} \sqrt{{dk_j \over dH}}\Bigg|_{H=E} 
<  u_i \left| V \right| u_j^\pm >
\ea
The unitarity condition $S^+ S= I$
provides a check on the 
precision of the results, and was fulfilled within a typical error $10^{-10}$.
 
We have considered the dimensionless residual conductance
\ba
g=\frac{2 \pi \hbar}{e^2}  {\cal{G}} 
\ea
and investigated the scaling properties of the geometric  average
${g_{av}} = \exp{ \overline{\left< \ln{g} \right>} }
$ in the regime $1 \ll b \ll L$, where $L$ is the length of the 
disordered sample. The brackets $\left< \ldots  \right> $ 
mean here average over different realizations
 of the potential ($20000$ realizations in 
our computations) while the bar means average over
the energy range $(-1,1)$. As expected, ${g_{av}} $ 
turned out to depend on $b$ and 
$L$ only through the localization parameter $x=\frac{b^2}{L}$ (Fig.1),
where the physical meaning of $b$ and $b^2$ is that they are proportional
respectively to the mean free path and to the localization length.
For $x\gg 1$, i.e. in the delocalized regime, this dependence has the 
form ${g_{av}}\propto x\propto L^{-1}$, which is
the "ohmic" behavior (Fig. 2). Instead, in the localized regime $x \ll 1$, 
${ g_{av}}  \approx \exp( -c/x)$ (Fig. 3).

Our results can be summarized by the following interpolating law,
 valid in all regimes: 
\ba \label{ours}
{g_{av}} = \frac{a+b\exp{(-d/x)}}{\exp{(c/x)} -1}
\ea
where $c\approx 2.70$ and $d\approx 0.55$ (obtained from the
fitting in the 
localized region) and $a\approx 6.58$ and $b\approx -5.53$ 
(obtained from the fitting in the metallic region).
 If the conductance is computed 
without performing the energy average, a similar scaling law is obtained, 
though with slightly different constants (curves $\bf a$ and $\bf b$ in Fig. 4).

We now turn to the Thouless approach.
In order to define the  (dimensionless) Thouless conductance, the BRM which 
describes the disordered sample has to be 'periodicized' as described in 
ref. \cite{Molinari}. The 'periodic' BRM thus obtained describes a ring-shaped 
conductor, and depends on a phase $\phi$ that has the meaning of an 
Aharonov-Bohm flux switched on through the ring. The disorder-averaged 
Thouless conductance is
\ba \label{Thouless}
{\cal{K}} = \frac{1}{\Delta} \left< \left\vert \frac{d^2 E}{d \phi^2} 
\right\vert
\Bigg|_{\phi=0} \right>
\ea
where $\Delta$ is the mean level spacing, and $\langle \ldots \rangle$
 denotes statistical averaging.
Although the scaling behavior of $\cal{K}$ was 
described in ref. \cite{Molinari} by an interpolating law different from 
 Eq. (\ref{ours}),
the latter law yields a slightly better fit with the parameters
$a'\approx 0.11$, $b'\approx -3.1_{  } \cdot
    10^{-2}$, $c'\approx 1.56$ and
$d'\approx 0.99$. Therefore 
Eq. (\ref{ours}) can be used to describe the scaling of both the
 Landauer and the Thouless conductance, albeit with different parameters.

A key problem  in comparing the Thouless and the Landauer conductance is 
connected with the matching of the "free" and of the "disordered" part of 
the Hamiltonian, i.e., with the effect of "contacts", which was discussed 
in refs.\cite{Weid}. If the arbitrary scale factor $\alpha$ appearing 
in the disordered part of our Hamiltonian ${\cal{H}}$ is varied, 
 while keeping the free Hamiltonian  ${\cal{H}}_0$
unchanged, the Thouless 
conductance is not affected, nor is the value of the 
localization length in the 'infinite' sample. The Landauer conductance, 
instead, is to some extent modified. It becomes very small both 
at small and at large values of $\alpha$, and it has a relatively broad maximum 
around $\alpha=1$. A value of $\alpha$ close to $1$ is also found on requiring 
that the local density of states for the "free" dynamics be the same as the 
density of interacting states in the sample (having computed the latter from 
the interacting Green function, in a standard way). Our choice of 
$\alpha=1$, which maximizes the transmission, is very similar to the 
"matching wire" prescription, 
suggested by Economou and Soukoulis \cite{eco}.

The comparison between Landauer and Thouless conductance, 
computed as described above, 
is shown in Fig.4.
In particular, in the metallic regime ($x \gg 1$) we have 
${g_{av}} \approx (7.5 \pm 0.4) {K_{av}} +0.8$,
that is, the two conductances are proportional in 
the diffusive regime, where
(\ref{Thouless}) is assumed to be valid. This 
difference cannot be removed by using of a different expression
for the Landauer formula, as in \cite{Buttiger}.
The meaning of the coefficent of proportionality $\approx 7.5 \pm 0.4$ is not 
clear to us \cite{lmont}. It is of course possible that this coefficient
 becomes closer to $1$ in the 
limit $b\to\infty$, $L\to\infty$, $b^2/L=const.$; 
this we were unable to check 
due to obvious numerical limitations, but, in the parameter range we were able 
to explore, the variation of this  factor is quite slight.  

In the localized regime ($x \ll 1$) we get 
$\ln({g_{av}}) \approx \beta \ln({K_{av}})+5.66$, that is to say, the 
Landauer conductance is proportional to $K_{av}^{\beta}$, 
with $\beta=1.7$. 
The error in this numerical estimate is of order $0.1$, due to the difficulty 
in computing
curvatures in the localized regime.
In this connection, we have also studied two different tridiagonal 
models: a BRM with
$b=1$ and a standard Anderson model with Gaussian disorder on the main diagonal.
Comparing Landauer conductances and curvatures on these two models we got 
$\beta_{BRM}=2.08 \pm 0.04$ and $\beta_{And}=2.09 \pm 0.01$ 
\cite{Lucasugg}. It seems 
therefore reasonable \cite{Eric}
to conjecture $\beta=2$.

Let us now discuss the fluctuation properties of the conductance.
First, we have found that in the localized region the following relation
holds: 
$Var(\ln{g}) = -\kappa \left< \ln{g} \right> + \gamma$  with $\kappa=2.00
 \pm 0.02$,
and $\gamma=-4.3 \pm 0.5$ (Fig.5). 
This has to be compared with the variance of the curvature, which in 
\cite{Molinari} was found to obey  
a similar law, but with $\kappa \approx 1$.
This fact reflects the most serious difference between Landauer conductance
and curvature in the strongly localized regime.

We have also performed a detailed
analysis of the distribution $P(g)$ of conductance in  
different regimes. Our results are in perfect agreement with theoretical 
expectations for quasi-1D conductors\cite{Pichvar}, in both the localized and in
the metallic regimes. In particular, we have found that the shape of 
this distribution 
is controlled by the same scaling parameter $x$: it is log-normal 
in the localized region,  $x \ll 1$, and  normal in the
delocalized one, $x \gg 1$.  

Special attention was paid to universal conductance fluctuations in the
metallic regime. We have performed numerical simulations for large values
of the scaling parameter  ${b^2 \over L}$, which is proportional
to the ratio of the localization length (in the infinite sample) to the finite
 sample size. At the same time, we have taken the ratio  ${b \over L}$ small,
 that is, we have taken samples considerably larger than the mean free
path.
Our data for the variance of conductance versus the scaling parameter are 
summarized in Fig. $6$. For not very large values of  $ x \approx 5 $
the value of $Var(g)$ remains close to $ 1\over 8$, slightly different
from the expected value $2 \over 15$.
As discussed in \cite{Beenmy}, the value $1 \over 8$ is predicted by RMT,
while $2 \over 15$ follows from different theoretical approaches, based
either on diagrammatic calculations, or on a diffusion-equation
approach for transfer matrices; it is also obtained from RMT, if an
appropriate repulsion law is used for eigenvalues of the transfer matrix.
More detailed numerical studies, at larger values of $x \approx 10$, reveal
that $2 \over 15$ is slowly approached on increasing $x$. All our results
were carefully checked by using a very large number of realizations, and by 
implementing different random number generators.

Our data indicate that the value $2 \over 15$ is approached very slowly, in 
such a way that  actual data linger around $1 \over 8$ in a significant 
parameter range. That the approach to $2 \over 15$ can be relatively slow
has been pointed out in Ref. \cite{Weid}, where this effect was related
to the coupling with channels.

Finally we would like to remark that, since the velocity in the $i$th channel is  
$v_i=\frac{d H}{d k_i} \Big|_{H=E}$, in our model 
different channels have different 
velocities.
This raises the question, how does this fact affect our results. We have then 
modified the structure of the leads, in such a way that the dispersion law
gives the same velocity in all channels. For this
we have chosen  the free hamiltonian with elements
 $H_{ij}=\delta_{ij}+\delta_{i+b+1,j}+\delta_{i-b-1,j}$.
The data obtained in this way are practically
indistinguishable from those   
in Fig. 1.
Therefore the fluctuation properties of conductance appear to be independent of
the specific properties of the leads.

In conclusion, we have studied the Landauer conductance for thin disorderd
wires attached to perfect leads in 1D (or in quasi-1D geometry).
The model we have used is based on Band Random Matrices with the modifications
imposed by the presence of the leads. The results are compared with those found 
for the Thouless conductance both in localized and metallic 
regimes. In the localized regime our results reveal that the Landauer
conductance is proportional to the square of the Thouless conductance. In the
metallic regime they are proportional to each other.
Therefore,
our data confirm theoretical expectations for the fluctuation properties
of conductance in the metallic regime.

\begin{figure}
\caption{ Logarithm of the geometric
 average of the Landauer conductance 
${ {g_{av}} } = \exp{ \overline{ \left< \ln{(g)} \right> }}$ vs.
logarithm of the scaling parameter $x={b^2 \over L }$.
Numerical data correspond to different values of $L$ and $b$ in the
ranges $50 \leq L \leq 1500$, $7 \leq b \leq 80$. 
Triangles are for the dispersion law ($2$), circles for a dispersion
law giving the same velocity in all channels.
The full line gives the 
fitting law eq.($8$). }
\end{figure}

\begin{figure}
\caption{ Geometric average of Landauer conductance vs.
the scaling parameter ${b^2 \over L}$ in the delocalized regime ($b^2 \gg L$).
 The linear fit shows the Ohmic
behavior of Landauer conductance:
$ {g_{av}}=\exp{\overline{ \langle 
\ln(g) \rangle}} = A {b^2 \over L} + B$, with $A=0.44 \pm 10^{-2}
$,
 $B=0.6 \pm 0.1$. }
\end{figure}

\begin{figure}
\caption{ $\ln{( {g_{av}} )}$ vs.
the inverse of the scaling parameter in the localized regime ($b^2 \ll L$).
 The linear fit shows the exponential
decrease of Landauer conductance due to
localization:  $\overline{ 
\langle \ln(g) \rangle}
= -{C {L \over b^2}} + D $, with $C=2.70 \pm 4 \cdot 
10^{-2}$, $D=1.8 \pm 0.2$.}
\end{figure}

\begin{figure}
\caption{ Comparison between the fittings of Landauer conductance,
computed with (curve $a$) and without (curve $b$) 
an energy average in the interval 
$(-1,1)$. Notice that the effect of
the energy average is very small. 
Curve c gives the fit of
Thouless conductance computed in ref.$[8]$. }
\end{figure}

\begin{figure}
\caption{ Variance of the logarithm of Landauer conductance $g$ vs. 
$< \ln (g) > $. The numerical fit shows that in the localized
regime $Var( \ln{(g)} ) = - \kappa \langle \ln{(g)} \rangle + \gamma$,
with $\kappa = 2.00 \pm 0.02$ and $\gamma = -4.3 \pm 0.5$.}
\end{figure}

\begin{figure}
\caption{ Variance of the logarithm of Landauer conductance vs. 
logarithm of the scaling parameter: the plateau in which $Var(g)$ is 
independent of $b$ corresponds to the regime of UCF, before the onset of the 
ballistic regime.
The theoretical expected value $2 \over 15$ is 
reached only for $x \approx 10$, while for
$x \approx 5$ data linger around $1 \over 8$ (upper and lower horizontal 
lines).}
\end{figure}


\begin{thebibliography}{10}


\bibitem{Thouless72}
D.Thouless J.T.Edwards.
\newblock {\it J.Phys.}, {\bf C5}, (1972) 807.

\bibitem{Thouless74}
D.Thouless.
\newblock {\it Phys.Rep.}, {\bf 13}, (1974) 93.

\bibitem{landauer57}
R. Landauer.
\newblock {\it I. B. M. Res. Dev.}, {\bf 1}, (1957).

\bibitem{landauer70}
R. Landauer.
\newblock {\it Philos. Mag.}, {\bf 21}, (1970) 873.
                                                  
\bibitem{fish}
D.S.Fisher and P.A.Lee.
\newblock {\it Phys. Rev.}, {\bf B23}, (1981) 6851.  

\bibitem{akkermans}
E.Akkermans and G.Montambaux.
\newblock {\it Phys.Rev.Lett.}, {\bf 68}, (1992) 642. 

\bibitem{fyod}
Y.V.Fyodorov and A.D.Mirlin.
\newblock {\it Int.J.Mod.Phys.}, {\bf 8}, (1994) 3795.

\bibitem{Buttiger}
M.B{\"{u}}ttiker, Y.Imry, R.Landauer, and S.Pinhas.
\newblock {\it Phys. Rev.}, {\bf B31}, (1985) 6207.             

\bibitem{Molinari}
G.Casati, I.Guarneri, F.M.Izrailev, L.Molinari, and K.\.{Z}yczkowski.
\newblock {\it Phys.Rev.Lett.}, {\bf 72}, (1994)  2697.


\bibitem{Weid}
S. Iida, H. A. Weidenm\"{u}ller and J.A.Zuk.
\newblock {\it Ann. Phys. (NY)}, {\bf 200}, (1990) 219.
                                                     
\bibitem{eco}
E.N.Economou and C.M.Soukoulis.
\newblock {\it Phys. Rev. Lett.}, {\bf 43}, (1981) 618.
                                                                  
\bibitem{lmont} After the conclusion of this work, we became aware of the 
related work: \\ 
\newblock D.Braun, E.Hofstetter, G.Montambaux and A.MacKinnon.
\newblock {\it preprint} cond-mat/9611059 

\bibitem{Lucasugg} These results were obtained computing curvatures
by means of a formula
valid only for tridiagonal models,which was pointed out to us by
Luca Molinari.                                                  

\bibitem{Eric} An argument to the effect that the dissipative conductance,
in the deeply localized regime, should be proportional to the square of the cur`
is based on the fact that in the Fourier expansion of energy levels with
respect to the phase $\phi$ only the lowest harmonics survive. This
argument was shown to us by Eric Akkermans.                         


\bibitem{Pichvar}
J.L.Pichard, N.Zanon, Y.Imry, and A.D.Stone.
\newblock {\it J.Phys.(France)}, {\bf 51}, (1990) 587.

\bibitem{Beenmy}
C.W.J. Beenakker.
\newblock {\it Mod. Phys. Lett.}, {\bf B8}, (1994) 469.

\end{thebibliography}
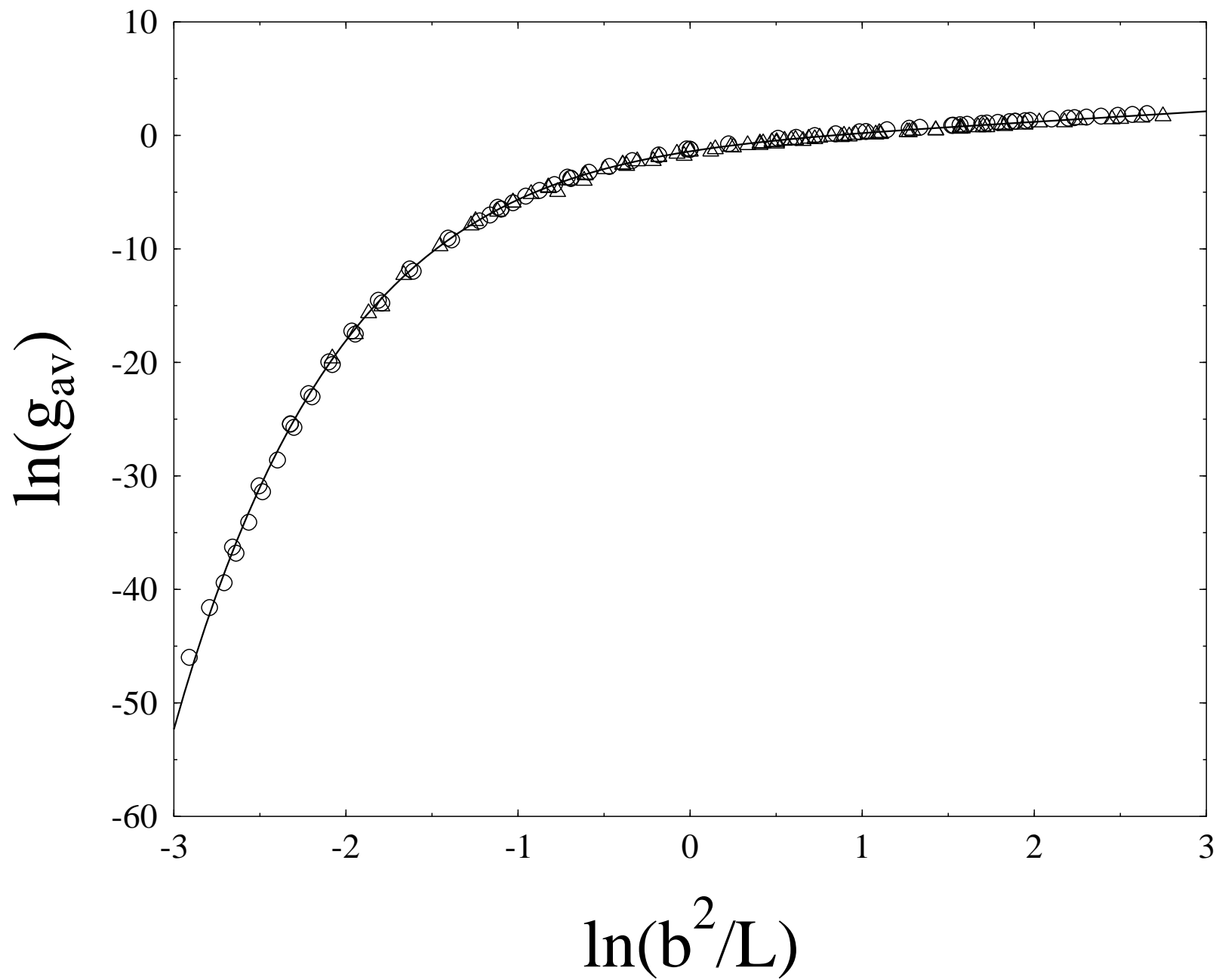
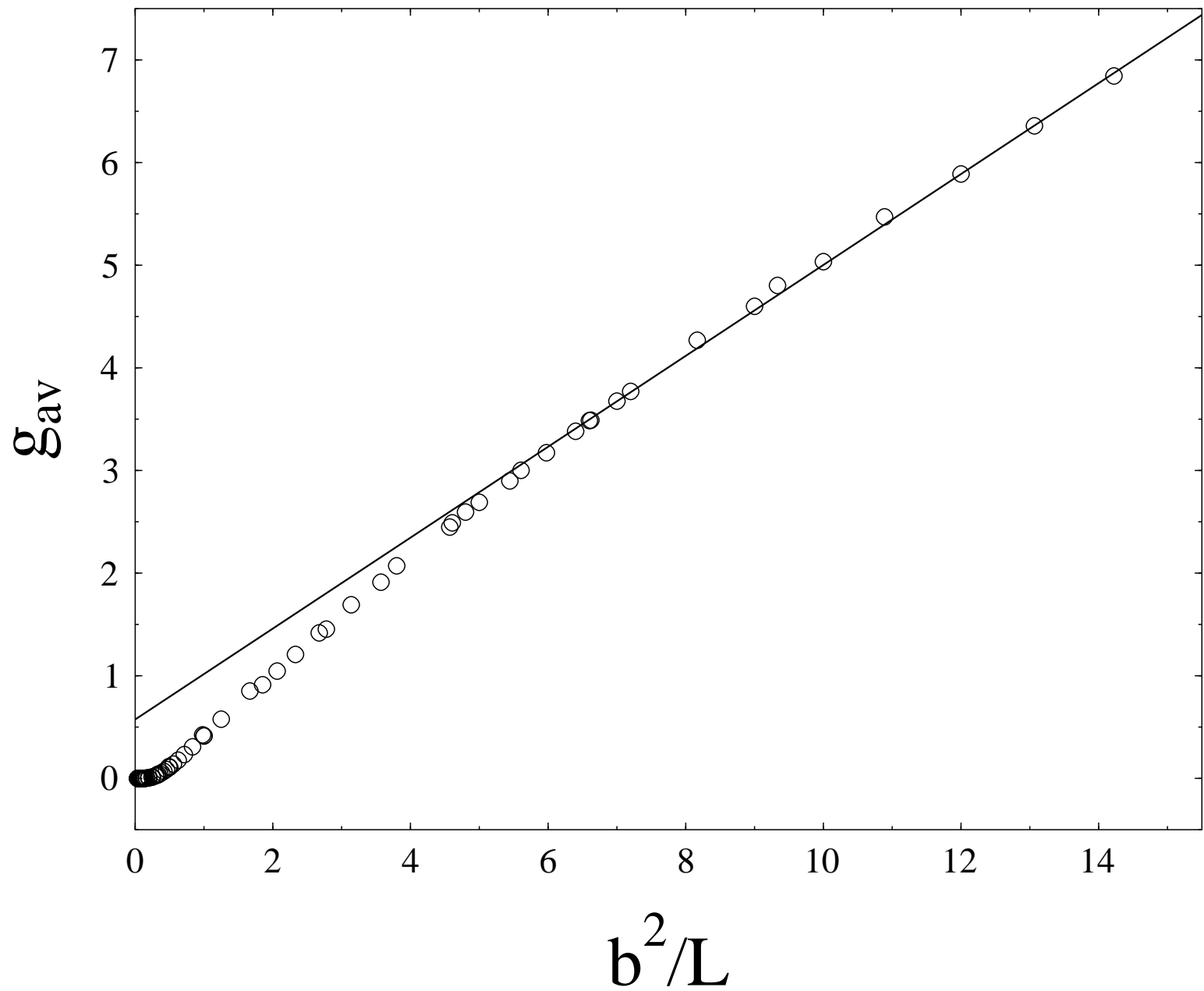
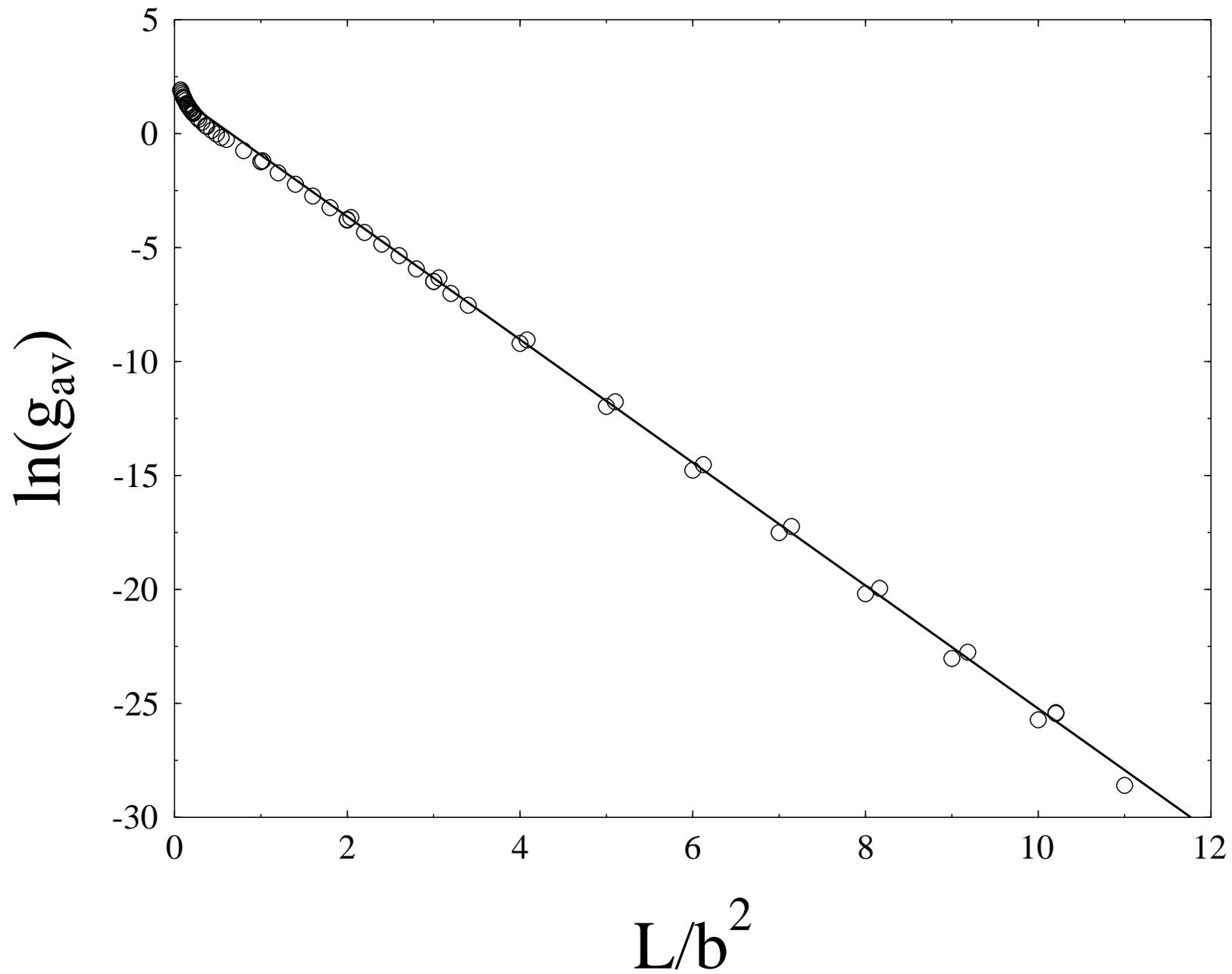
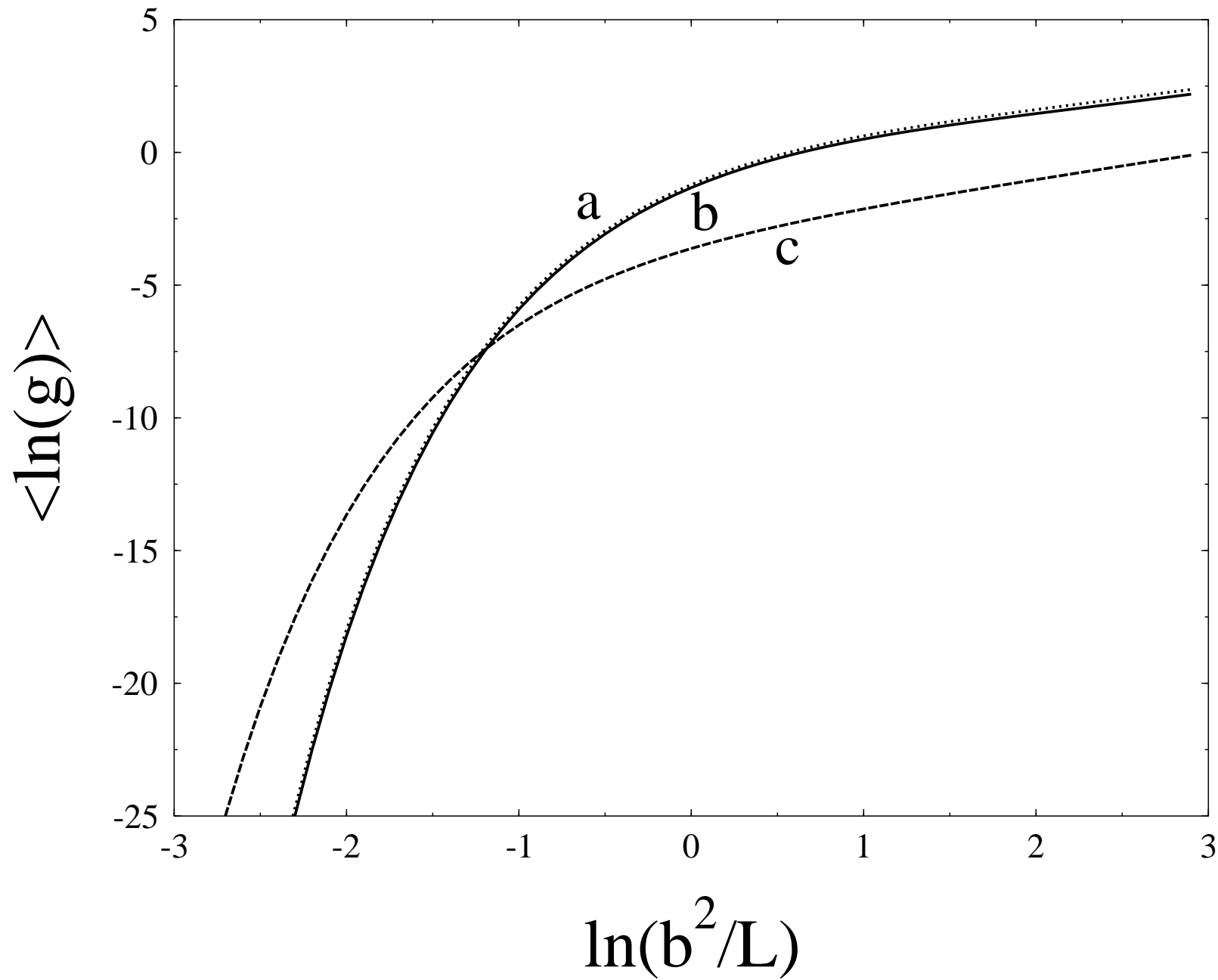
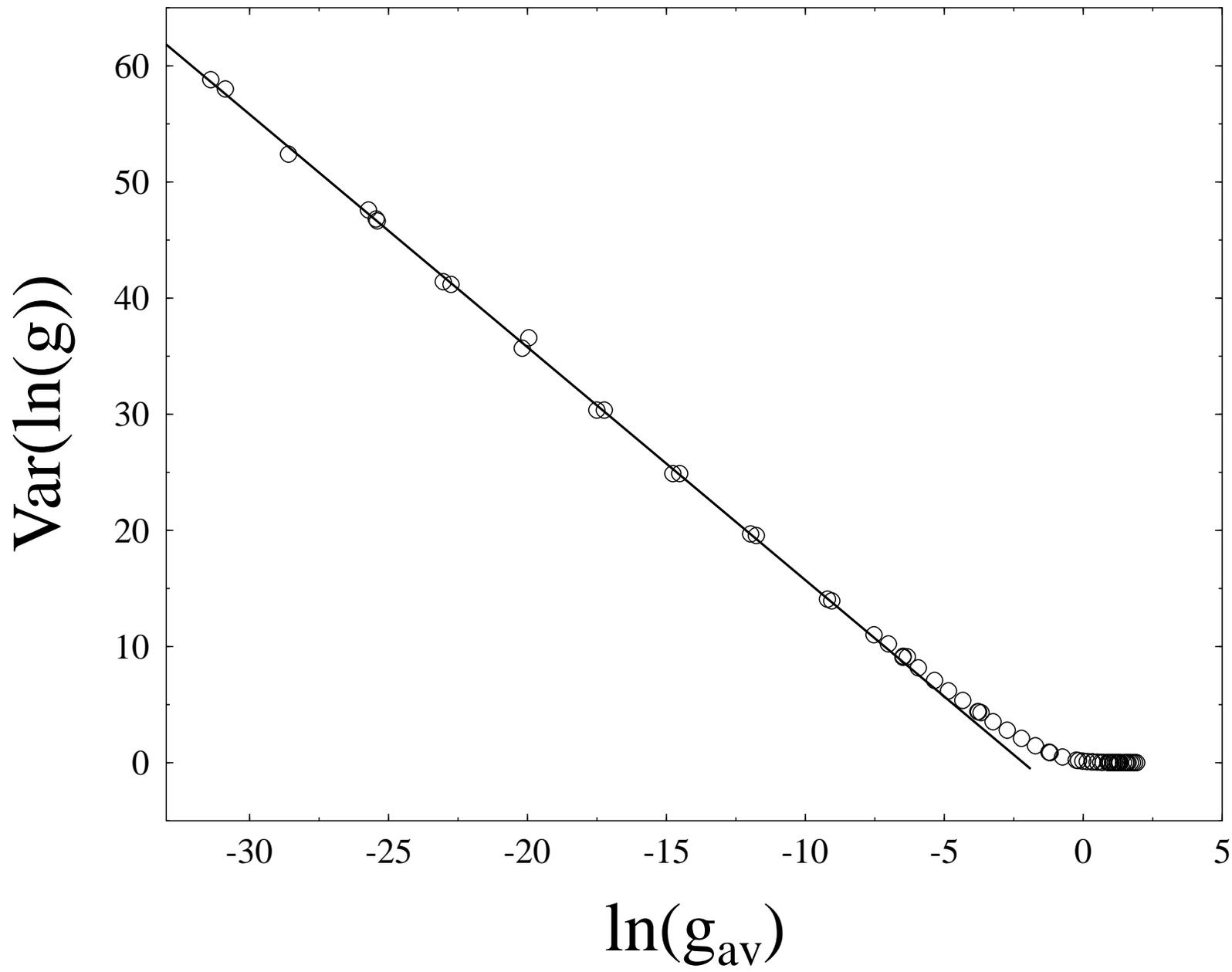
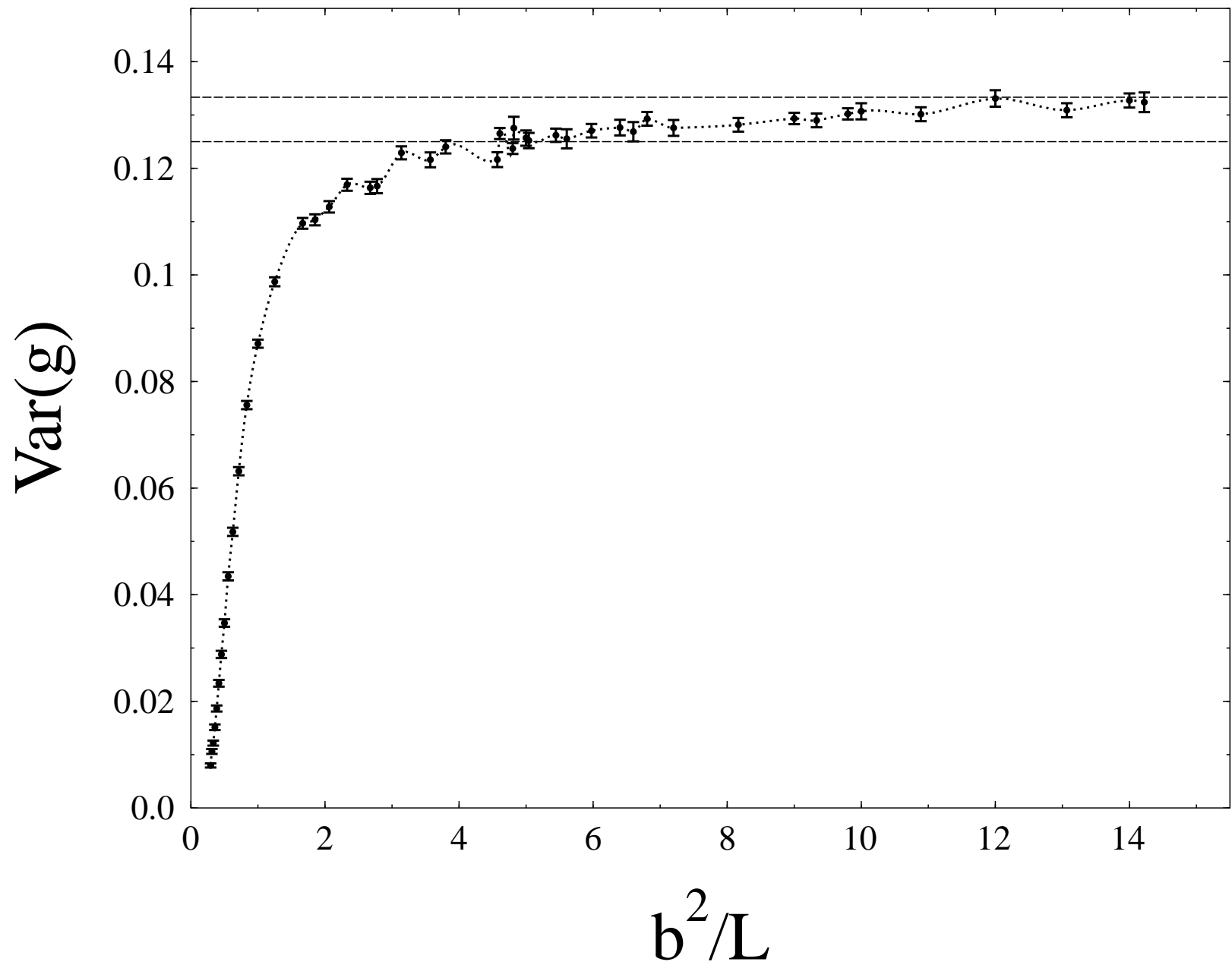
\end{document}